\documentclass[conference,letterpaper,10pt]{IEEEtran}
\IEEEoverridecommandlockouts
\usepackage{newtxtext,newtxmath}

\usepackage{cite}
\usepackage{enumitem}
\usepackage{amsmath,amsfonts}
\usepackage{algorithmic}
\usepackage{graphicx}
\usepackage{textcomp}
\usepackage{xcolor}
\def\BibTeX{{\rm B\kern-.05em{\sc i\kern-.025em b}\kern-.08em
    T\kern-.1667em\lower.7ex\hbox{E}\kern-.125emX}}
\begin{document}

\makeatletter
\newcommand{\linebreakand}{%
  \end{@IEEEauthorhalign}
  \hfill\mbox{}\par
  \mbox{}\hfill\begin{@IEEEauthorhalign}
}
\makeatother

\title{Beyond RSRP: An Exploration of Alternative Metrics for Beam Management in an Urban Micro 5G NR Scenario\\
\thanks{This research was supported by the U.S. Department of Commerce’s National Telecommunications and Information Administration (NTIA) under the Public Wireless Supply Chain Innovation Fund Grant Program (Award 24-60-IF2415: ASPEN - Advanced Signal Processing Enhancement for Next- Generation Open Radio Units), administered by the National Institute of Standards and Technology.}
}

\author{
    \IEEEauthorblockN{
         Saakshi Gupta\IEEEauthorrefmark{1},
         Tarun S. Cousik\IEEEauthorrefmark{1},
         Pratheek Upadhyaya\IEEEauthorrefmark{1},
         Daniel J. Jakubisin\IEEEauthorrefmark{1}\IEEEauthorrefmark{2},
         Nishith Tripathi\IEEEauthorrefmark{1},
         Jeffery H. Reed\IEEEauthorrefmark{1}
    }

    \IEEEauthorblockA{
        \IEEEauthorrefmark{1}Wireless $@$ Virginia Tech, Bradley Department of ECE, Virginia Tech, Blacksburg, VA USA
    }

    \IEEEauthorblockA{
        \IEEEauthorrefmark{2}Hume Center for National Security and Technology, National Security Institute, Virginia Tech, Blacksburg, VA USA
    }

    \IEEEauthorblockA{
        gsaakshi@vt.edu,
        tarunsc@vt.edu,
        pratheek@vt.edu,
        djj@vt.edu,
        nishith@vt.edu,
        reedjh@vt.edu
    }
}

\maketitle

\begin{abstract}
Reference Signal Received Power (RSRP) is widely used for beam selection in 5G NR as it is an easy indicator of path loss, but is not interference-aware, limiting its effectiveness in dense deployments. In this paper, we explore RSRP, Reference Signal Received Quality (RSRQ), and Signal to Interference-plus-Noise Ratio (SINR) as beam selection metrics alongside three UE-gNB association strategies in a 3GPP-compliant Urban Micro multi-cell simulation, benchmarked against Regularized Zero-Forcing (RZF) full-Channel State Information (CSI) upper bounds. Using a 3GPP-compliant Urban Micro multi-cell simulation benchmarked against RZF full-CSI upper bounds, we demonstrate two findings: directional RSRQ, despite its inter-cell interference awareness, is the worst-performing strategy in symmetric interference, and sequential association outperforms power-optimal Hungarian assignment. We also establish that the residual performance gap to full-CSI precoding stems from codebook quantization rather than metric imperfection, identifying adaptive codebook design as a mechanism to enhance performance.
\end{abstract}


\section{Introduction}
5G New Radio (NR) enables unrivaled data rates, low latency, and massive connectivity across mmWave and sub-6 GHz bands, with beam management central to realizing these gains \cite{tripathi20195g}. Beam management is the set of Layer 1 or PHY (L1) and Layer 2 or MAC (L2) procedures that establish and maintain an optimal beam pair between a ground node-B (gNB) and User Equipment (UE) \cite{mathworks_white_paper}, and this paper focuses on the Urban Micro deployment scenario.\\
Beam selection has traditionally relied on RSRP, but RSRP measures total received power, including interference and noise \cite{etsits138215}, making it poorly suited for interference-limited deployments. 5G NR also defines RSRQ and SINR, which account for interference and directly quantify link quality, yet in-depth studies comparing these metrics for beam management in realistic multi-cell scenarios remain limited in the literature.\\
Assuming beam pairs have been set-up via initial cell search and Master Information Block (MIB) synchronization, beam management proceeds through five stages:
\begin{itemize}
\item \textbf{Beam Sweeping:} the gNB transmits beams in all directions; each UE selects the best available beam during initial access.
\item \textbf{Beam Measurement and Determination:} the UE measures each beam pair and identifies the strongest based on RSRP.
\item \textbf{Beam Reporting:} the UE reports the best beam to the gNB via a measurement report.
\item \textbf{Beam Recovery:} if the active beam pair fails due to poor channel conditions, the UE reports a new candidate beam.
\item \textbf{Beam Switching:} if a better beam pair is found, the UE switches beams through L1/L2 mobility procedures.
\end{itemize}
This paper examines the Beam Measurement stage using Synchronization Signals (SS) reference signals (RS) rather than CSI-RS, as SS signals are cell-wide broadcasts suited to initial access and mobility. We evaluate whether incorporating interference-aware metrics (RSRQ, SINR) into beam selection improves over RSRP, framing this as implicit interference excision within the L1 signal strength maximization problem. Section \ref{sec:limitations} discusses the shortcomings of RSRP, and Section \ref{sec:alt metrics} presents the alternative metrics considered. Section \ref{sec:system model and analysis} addresses the system model and defines the performance metrics for this study, while Section \ref{sec:results} and \ref{sec:conclusion} discuss the results and conclude the study, respectively.

\section{Different Approaches to Beam Management}\label{sec:limitations}
Beam management research spans several directions, of which initial access, alternative performance metrics, and machine learning approaches are most relevant.\\
For initial access and beam alignment, \cite{khalili} formulates the problem as minimizing uncertainty in angles of departure (AoDs) across multi-user feedback, providing information-theoretic bounds on optimal performance. The authors of \cite{dreifuerst_hierarchical_2024} address extreme MIMO systems via X-BM, which unifies beamforming and CSI acquisition through hierarchical codebook design learned from beam space representations, achieving an 8 dB improvement in initial access and gains in effective spectral efficiency.\\
Beyond RSRP, recent work evaluates beam management using a broader set of metrics. Spectral efficiency is explicitly optimized in \cite{Narengerile2021DeepRL} via deep Reinforcement Learning and reported for extreme MIMO in \cite{dreifuerst_hierarchical_2024}. Latency reductions of $50\%$ \cite{arvinte}, $3\times$ \cite{heng}, and $7×\times$ via DeepBeam \cite{polese} over standard beam sweeping have been demonstrated. Beam prediction accuracy of $98.27\%$ \cite{zhang2025channelchartingbasedfast}, $94.8\%$ \cite{selvaperumal}, and $40.58\%$ over baselines \cite{jalali} is reported. Grid-free beam alignment achieves $10$ to $15$ dB better average SNR than the conventional codebook search \cite{heng_grid_free_mimo}.
CNNs are the dominant ML architecture: \cite{jalali} processes RSRP measurements to predict beam pairs, and DeepBeam \cite{polese} infers beam information directly from waveforms without explicit coordination. End-to-end learning jointly optimizes probing codebooks and predictors \cite{heng} and hierarchical codebooks \cite{dreifuerst2023hierarchicalmlcodebookdesign}. Reinforcement learning frames beam selection as sequential decision-making; \cite{Narengerile2021DeepRL} adapts to channel conditions by switching beam training strategies and controlling RF chain activation to balance energy efficiency and spectral efficiency for millimeter wave systems.

\subsection{Limitations of RSRP}
RSRP measures received signal power averaged over reference signal resource elements \cite{etsits138215}, making it straightforward to compute and report. In 5G NR, SS-RSRP is specifically computed from SSS-carrying resource elements within each SS Block (SSB) transmission. Since the SSS is swept across multiple beams, one SS-RSRP value is obtained per beam, enabling beam-specific signal strength comparison. However, because it captures only signal strength and carries no information about interference, selecting a beam that maximizes RSRP does not generally maximize post-equalization SINR or user throughput, particularly in interference-limited deployments \cite{giordani_tutorial_2019}. In dense multi-cell networks, a beam pointing toward the strongest signal may simultaneously align with a strong interference corridor, causing RSRP-optimal selection to be SINR-suboptimal \cite{tr_138_901}.

\section{Exploring Alternate Metrics for Beam Management}\label{sec:alt metrics}
\subsection{Alternate Metrics}
\subsubsection{RSSI}
RSSI is a received power quantity that represents the total signal strength observed at the receiver's antenna, measured in dBm across the occupied bandwidth. While simple to acquire, it reflects only raw signal strength and cannot distinguish the desired signal from noise or interference, making it unreliable in multipath or high-interference conditions.
\subsubsection{RSRQ}
SS-RSRQ is the ratio $\left( N_{\text{RB}} \cdot \frac{\text{SS-RSRP}}{\text{RSSI}} \right)$, normalizing desired signal power against total received power over $N_{\text{RB}}$ resource blocks \cite{etsits138215}. This offers a more interference-aware view of beam quality than RSRP, though it is insufficient on its own to determine the optimal beam.
\subsubsection{SINR}
SS-SINR is the ratio of the linear average power of SSS-carrying resource elements to the average noise and interference power over the same resources, providing the most direct indicator of beam demodulation quality and achievable throughput. Since its numerator captures desired signal power and its denominator captures interference plus noise, SS-SINR is the closest standardized proxy for Shannon capacity, making it the most direct indicator of achievable throughput per beam. In 5G NR, the SSS is transmitted across multiple beams via SSB sweeps, yielding one SS-SINR measurement per beam during initial access.
\subsection{Phase- and GNSS-based Metrics}
Other candidate metrics include Global Navigation Satellite System (GNSS)-based measurements (frame timing, code measurements) and phase-based (Angle of Arrival (AOA), Angle of Departure (AOD), etc.). Phase-based metrics can, in principle, inform beam direction selection directly, since AOA and AOD estimates characterize the dominant propagation directions between UE and gNB, which map naturally onto codebook beam indices. On the other hand, GNSS measurements are suited primarily to satellite positioning\cite{Bnilam2025TwoStageBT}, while phase-based metrics require computationally intensive per-antenna channel estimation for active arrays, limiting their practical usefulness \cite{phase_noise_beamforming}.

\section{System Setup and Performance Analysis}\label{sec:system model and analysis}


\subsection{System Model}
We consider a 7-cell downlink simulation with 3 sectors per cell arranged on a hexagonal grid with an inter-site distance of $200$ m and gNB height of $10$ m, consistent with the 3GPP UMi street canyon model \cite{tr_138_901}. Each sector is served by one gNB, giving a total of 21 gNBs and 105 UEs (5 per sector). UE positions follow a uniformly random distribution subject to a minimum separation of $10$ m from the gNB, sector boundary, and angular constraints; UE heights range from $1.5$ to $22.5$ m with a $80\%$ indoor probability \cite{tr_138_901}. The resulting topology is shown in Fig. \ref{fig:3d_topology}: black markers denote gNBs, colored arrows show UE orientations, and UEs are classified as indoor (blue, Outdoor-to-Indoor, O2I low-loss path loss), outdoor Line of Sight, LoS (red), or outdoor Non-Line of Sight, NLoS (yellow), labeled $0, \dots, 104$. \\
\begin{figure}[tbp!]
    \centering
    \includegraphics[width=\columnwidth]{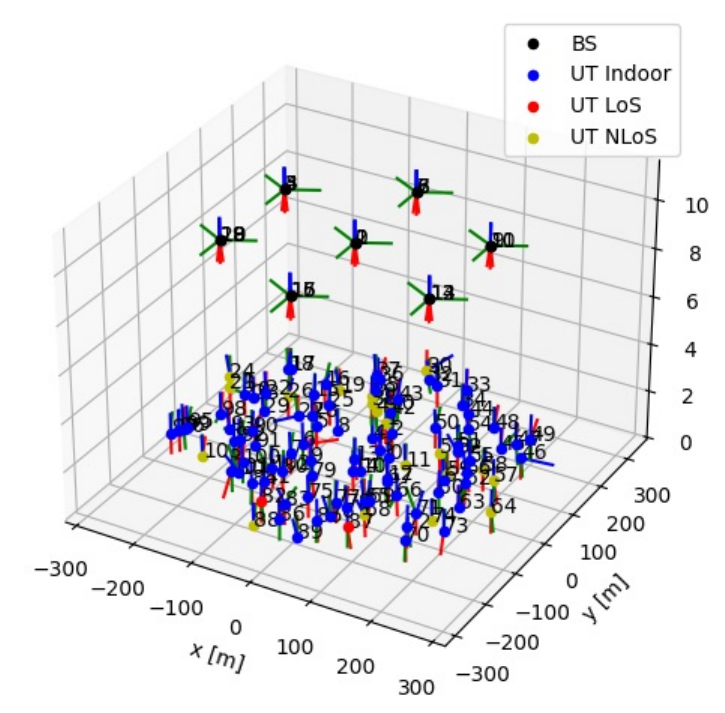}
    \caption{3GPP-compliant UMi multicell topology for a
             single simulation drop. BS is representative of gNBs; UTs are representative of UEs.}
    \label{fig:3d_topology}
\end{figure}
The system operates at $3.5$ GHz with $15$ kHz subcarrier spacing, $48$ active subcarriers ($720$ kHz bandwidth \cite{3gpp.38.104}), $14$ symbols per slot, and $1$ ms slots. The gNB and UE transmit power levels are $44$ dBm and $23$ dBm \cite{jain2024downlinkpowercontrolbased} respectively, and thermal noise is computed as $N_0 = k_B T \Delta f$, where $k_B$ is Boltzmann's constant, $T=294$ K, and $\Delta f = 15$ kHz is the subcarrier spacing. \\
In the context of the Sionna framework and this paper, one sector corresponds to one Base Station  (i.e., 5G gNB), and in practice, a Base Station typically covers three $120$ degree sectors to provide full 360 degree coverage. Each gNB has a $4\times4$ dual cross-polarized (VH) panel array ($32$ antenna ports, 3GPP 38.901 pattern) and each UE has a single vertically polarized omnidirectional antenna. Path loss, shadow fading, and the O2I low-loss model are applied \cite{tr_138_901}, with Large Scale Pathloss (LSPs) regenerated independently per drop. \\
UEs are associated with gNBs via large-scale fading power extracted from a single-step channel impulse response, solved as a constrained assignment using the Hungarian algorithm (5 UEs per gNB) \cite{hungarian}. The Hungarian algorithm finds the optimal one-to-one assignment between UEs and gNBs that maximizes on total received power in polynomial time \cite{obi2019hybrid}. The stream management object encoding this association governs precoder and equalizer design, and the OFDM channel matrix is generated from the same LSP realization to ensure consistency.
In the downlink scenario, each gNB applies RZF precoding:
\begin{equation}
    \mathbf{W} = \mathbf{H}^H(\mathbf{H}\mathbf{H}^H + \alpha \mathbf{I})^{-1}
\end{equation}
where $\mathbf{H}$ is the stacked channel matrix of all served UEs, and $\alpha$ is the regularization parameter set to the per-subcarrier noise power $N_0$. This regularization prevents noise amplification in ill-conditioned channels while spatially suppressing inter-user interference. An LMMSE equalizer at each UE recovers the desired stream, with post-equalization SINR accounting for residual leakage. Results are aggregated over 1000 Monte Carlo drops, yielding $105,000$ UE observations (Section \ref{sec:results}). Dropping UEs at independent uniformly random positions in each simulation drop implicitly captures the diversity of UE locations across mobility states, without explicitly modeling continuous UE trajectories.
\subsection{Association Strategies and Beam Selection Criteria}
This study decouples beam management into two design dimensions: \textit{Association Strategy} for how 5 UEs per sector are assigned to a gNB, and \textit{Beam Selection Criterion} for which codebook beam the gNB uses per UE.
\subsubsection{Association Strategies}
\begin{enumerate}[label=(\alph*)]
\item \textbf{Sequential (Baseline)}: UEs are assigned to gNBs based on their geographic location. The first $5$ UEs are assigned to the first gNB, the second $5$ UEs to the second gNB,  and so on.
\item \textbf{Hungarian RSRP}: The Hungarian algorithm maximizes total received power across all UE-gNB pairs with each gNB serving exactly 5 UEs, the standard 3GPP power-optimal baseline \cite{hungarian}.
\item \textbf{Hungarian SINR}: Each UE is assigned to the gNB offering the highest estimated link SINR rather than raw received power.
\end{enumerate}

\subsubsection{Beam Selection Criterion}
    Three criteria are applied on top of each association strategy:
    \begin{enumerate}[label=(\alph*)]
        \item \textbf{RSRP}: this is the standard 3GPP reporting metric where for each UE $u$ assigned to gNB $j$, the beam index $\hat{m}$ is selected as:
        \begin{equation}
            \hat{m} = \arg\max_{m} \left\{ p_{\text{tx}} \cdot \left| \bar{\mathbf{h}}_{u,j}^H \mathbf{w}_m \right|^2 \right\}
        \end{equation}
        where, 
        \begin{equation}
            \bar{\mathbf{h}}_{u,j}^H = \frac{1}{N_s N_f} \sum_{s,f} \mathbf{h}_{u,j,s,f}
        \end{equation}
        is the spatially averaged channel vector, $p_\text{tx}$ is the transmit power, $N_S = 14$ is the number of OFDM symbols, $N_f = 48$ is the number of subcarriers,  and $\mathbf{w}_m$ is the $m$-th codebook column. Since inter-cell interference enters pre-equalization SINR as a beam-independent scalar, maximizing RSRP and maximizing pre-equalization SINR select identical beams.
        \item \textbf{SLNR (Signal to Leakage plus Noise Ratio)}: maximizes the ratio of desired power to intra-cell leakage plus noise:
        \begin{equation}
            \hat{m} = \arg\max_{m} \left\{ \frac{p_{\text{tx}} \cdot \left|\bar{\mathbf{h}}_{u,j}^H \mathbf{w}_m\right|^2} {\sum \limits_{v \in \mathcal{S}_j \setminus \{u\}} p_{\text{tx}} \cdot \left|\bar{\mathbf{h}}_{v,j}^H \mathbf{w}_m\right|^2 + N_0 N_f} \right\}
        \end{equation}
        where $\mathcal{S}_j$ is the set of UEs served by gNB $j$, $u$ is the UE whose beam we are interested in, whereas, $v$ index runs through all \textit{other} UEs within the same sector. The denominator varies with $m$ penalizing beams that cause high intra-cell leakage.
        \item \textbf{Directional RSRQ}: This metric maximizes on the directional RSRQ:
        \begin{equation}
            \hat{m} = \arg\max_{m} \left\{ \frac{N_{\text{RB}} \cdot p_{\text{tx}} \left|\bar{\mathbf{h}}_{u,j}^H \mathbf{w}_m\right|^2} {\sum \limits_{k=1}^{K} p_{\text{tx}} \cdot \left| \bar{\mathbf{h}}_{u,k}^H \mathbf{w}_m \right|^2 + N_0 N_f} \right\}
        \end{equation}
        where the denominator sums projected power from all $K=21$ gNBs in direction $m$, penalizing beams aligned with inter-cell interference corridors.
    \end{enumerate}

\subsubsection{Codebook}
Using a 2-D Kronecker product, we define a 2-D DFT based codebook over the $4 \times 4$ panel of the gNB antenna array. Since the array is dual polarized, there are $4 \times4 \times 2 \times 2 \times 2 = 128$ beams that span both polarizations in both azimuth and elevation angles.
\subsubsection{Post-Equalization SINR}
The post-equalization SINR for each [association, beam metric] pair is calculated explicitly for that beam-domain model. Given the selected beam $\hat{m}_u$ for each UE $u$ and the spatial channel $\bar{\mathbf{H}}$, the received power at UE $u$ from gNB $j$ on codebook beam for UE $v$ is:
\begin{equation}
    r_{u,j,v} = p_{\text{tx}} \left|\bar{\mathbf{h}}_{u,j}^H \mathbf{w}_{\hat{m}_v}\right|^2
\end{equation}
The desired signal power for UE $u$ is $r_{u, j_u, u}$, where $j_u$ is its serving gNB. The post-equalization SINR is:
\begin{equation}
    \text{SINR}_u = \frac{r_{u, j_u, u}}{p \cdot \max\left(0,\sum \limits_{j,v} r_{u,j,v} - r_{u, j_u, u}\right) + N_0 N_f}
\end{equation}
where $p = 0.30$ is the scheduling probability applied to the interference term to model partial resource utilization. 
\subsubsection{Spectral Efficiency evaluation}
Once this scaling is applied identically in all seven codebook and RZF (Regularized Zero-Forcing) strategies, ensuring a consistent comparison. An RZF precoded multi-user MIMO systems outperforms Zero-Forcing (ZF) and Minimum Mean Squared Error (MMSE) corresponding precoded systems in terms of the bit error rate performance \cite{rzf_paper}, and hence serves as an upper bound to our analysis. Spectral efficiency (SE) is then calculated as  
\begin{equation}
    \text{SE}_u = \log_2(1 + \text{SINR}_u) \text{ bps/Hz}
\end{equation}
The seven strategies evaluated in the Monte Carlo process are: 
\begin{enumerate}[label=\roman*)]
    \item Sequential + RSRP beam,
    \item Hungarian RSRP + RSRP beam, 
    \item Hungarian SINR + RSRP beam,
    \item Hungarian SINR + SLNR beam,
    \item Hungarian SINR + Directional RSRQ beam,
    \item RZF Full CSI with Hungarian RSRP association, and 
    \item RZF Full CSI with Hungarian SINR association.
\end{enumerate}

\section{Results}\label{sec:results}
\subsection{Metric Distribution}
\begin{figure}[htbp]
    \centering
    \includegraphics[width=\columnwidth]{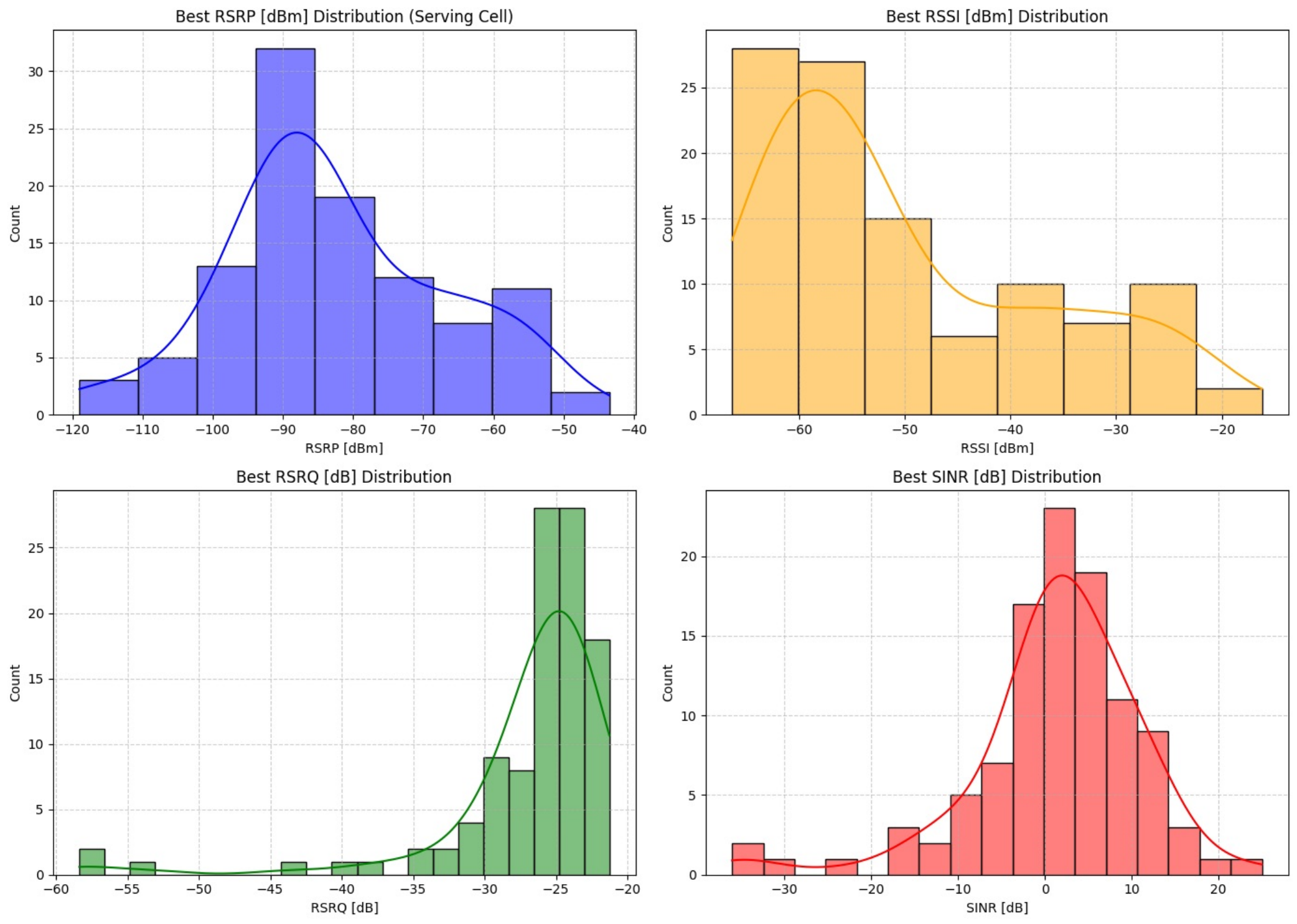}
    \caption{Empirical distributions of the four serving-cell radio
             metrics for all 105 UEs in a single representative simulation drop.}
    \label{fig:metric_distribution}
\end{figure}
The value of each metric that a UE can report needs to adhere to the 3GPP-defined reporting range. In Fig. \ref{fig:metric_distribution}, the empirical distributions of RSRP, RSSI, RSRQ, and post-equalization SINR for the serving link across all $105$ UEs in a representative single drop is to be observed. RSRP ranges from approximately $-115$ to $-40$ dBm, representing the diversity of indoor and outdoor propagation conditions. RSSI, which aggregates wideband power including inter-cell interference, shows a more compressed distribution because the symmetric hexagonal topology produces a roughly homogeneous interference floor and confirms that SINR carries the richest information for beam management decisions interference floor across all UEs. RSRQ is tightly clustered between $-40$ and $-20$ dB, tracking the wideband interference load. Post-equalization SINR spans $-20$ to $+30$ dB, showing the widest spread and carrying the richest spatial information for beam management. This range directly translates to the broadest spread in achievable spectral efficiency and confirms that SINR carries the richest information for beam management decisions. From these distributions, we can note that post-equalization SINR captures interference-aware information that can be useful for the beam-selection criteria, something the other power metrics are unable to.

\begin{figure}[htbp]
    \centering
    \includegraphics[width=\columnwidth]{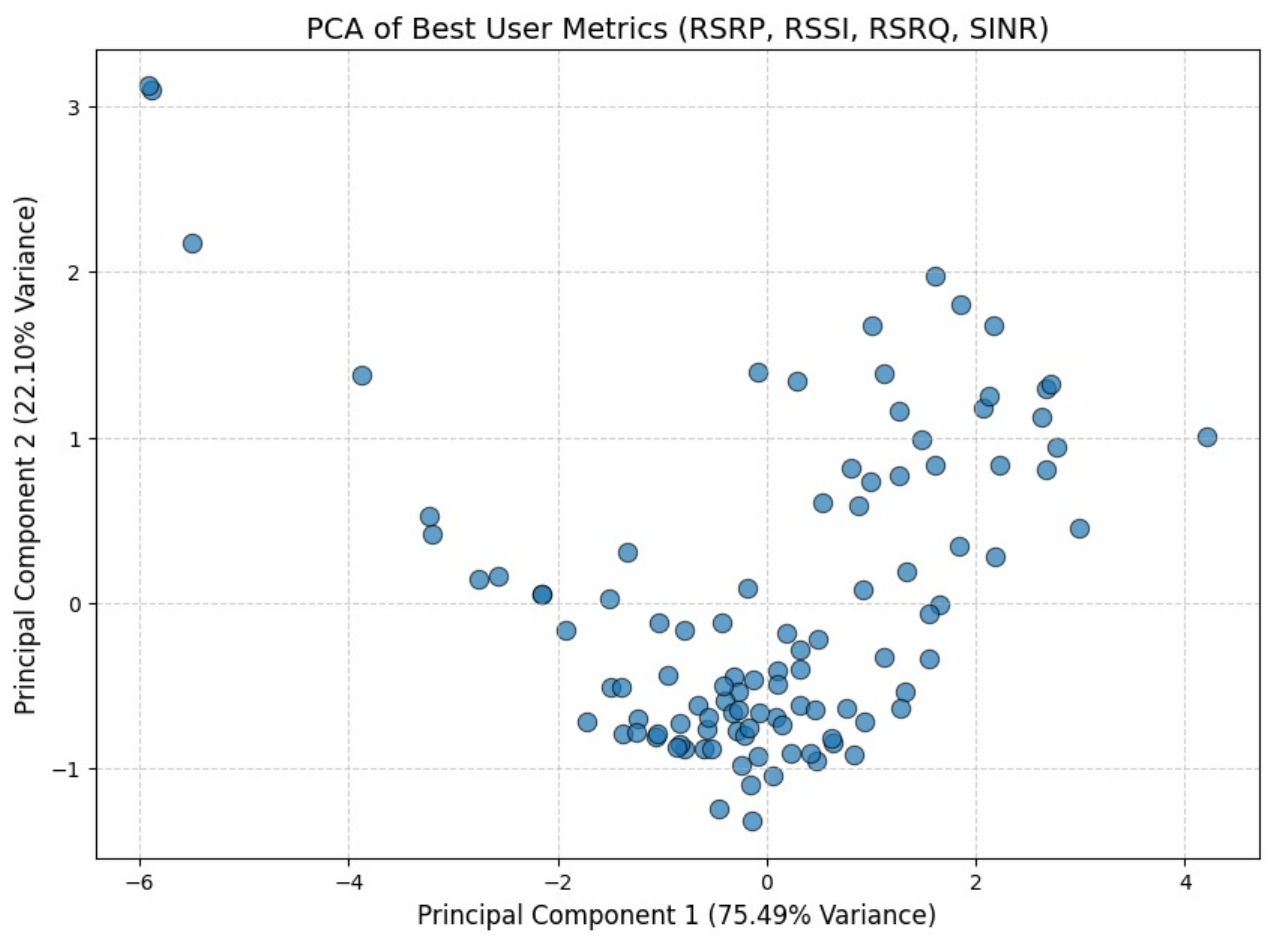}
    \caption{PCA projection of the four standardized serving-cell
             metrics (RSRP, RSSI, RSRQ, SINR) onto the first two
             principal components. Together, the two components explain 94.09\% of the total variance.}
    \label{fig:pca}
\end{figure}
\subsection{Principal Component Analysis}
PCA on the four standardized metrics (Fig. \ref{fig:pca}, Table \ref{tab1}) reveals two components accounting for 97.59\% of variance. PC1 (75.49\%) has near-equal positive loadings from RSRP (0.546), RSSI (0.462), RSRQ (0.466), and SINR (0.520), forming a general link-quality axis. PC2 (22.10\%) is dominated by opposing loadings from RSSI ($+0.634$) and RSRQ ($-0.590$), with SINR loading negatively ($-0.378$) and RSRP positively ($+0.327$). PC2 captures the interference load of a link: UEs with high total received power relative to their RSRQ and SINR  score positively, indicative of interference-dominated conditions, while UEs on cleaner links score negatively. The four metrics remain largely co-linear in this topology, with PC2 differentiating UEs by their interference environment rather than signal strength. The dominance of PC1 in this kind of symmetric deployment indicates that substituting one metric for another is only able to provide minimal additional information that can be used for beam management.
\begin{table}[htbp]
\caption{PCA Components with feature weights}
\begin{center}
\begin{tabular}{|p{0.5cm}|p{1.5cm}|p{1.5cm}|p{1.5cm}|p{1.5cm}|}
\hline
& \textbf{Best RSRP(dBm)}& \textbf{Best RSSI(dBm)}& \textbf{Best RSRQ(dB)}& \textbf{Best SINR(dB) }\\
\hline 
\textbf{PC1}& $0.546432$& $0.461932$& $0.466034$& $0.520427$  \\
\hline
\textbf{PC2}& $0.327119$& $0.634017$& $-0.590239$& $-0.377670$  \\
\hline
\end{tabular}
\label{tab1}
\end{center}
\end{table}
\begin{figure*}[htbp!]
    \centering
    \includegraphics[width=\textwidth]{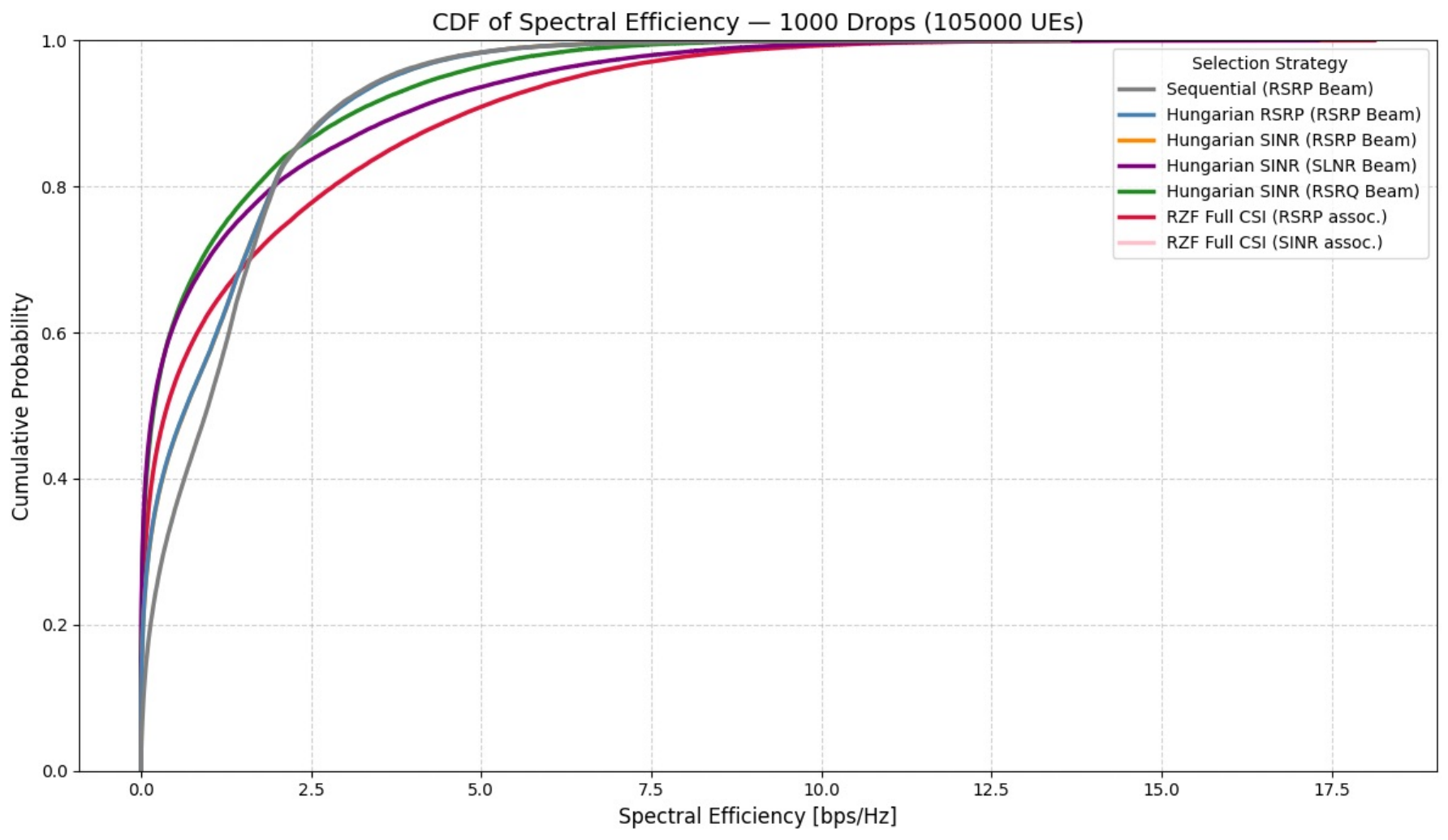}
    \caption{CDFs of spectral efficiencies applied over all selection strategies.}
    \label{fig:cdf se}
\end{figure*}
\subsection{Spectral Efficiency across Association and Beam Selection Criteria}
Fig. \ref{fig:cdf se} and Table \ref{tab2} summarize spectral efficiency for all seven strategies over $N_{\text{drops}} = 1000$ Monte Carlo drops. Sequential association ($1.23$ bps/Hz) outperforms both Hungarian strategies ($1.10$ bps/Hz each). Directional RSRQ ($0.95$ bps/Hz) is the worst-performing strategy: in a symmetric interference environment, beam-resolved interference avoidance selects geometrically poor beams. SLNR ($1.15$ bps/Hz) provides modest gains over RSRP-beam strategies. The near-identical RZF bounds ($1.49$ bps/Hz each) confirm that SINR-aware association provides negligible benefit under full-CSI precoding in the simulation setup, and the gap to the RZF ceiling points to codebook quantization as the primary performance bottleneck. Due to the symmetric nature of interference in this environment, the choice of beam selection strategy and/or association strategy to optimize beam management is mainly constrained by codebook quantization.
\begin{table}[htbp]
\caption{Mean Spectral Efficiency by Strategy}
\begin{center}
\begin{tabular}{|p{2.5cm}|p{1.5cm}|p{1.5cm}|p{1.5cm}|}
\hline
\textbf{Strategy}& \textbf{Association}& \textbf{Beam}& \textbf{Mean SE (in bps/Hz) }\\
\hline 
Baseline Sequential& Sequential& RSRP& $1.227689$  \\
\hline
Hungarian RSRP + RSRP& Hungarian RSRP& RSRP& $1.104723$  \\
\hline
Hungarian SINR + RSRP& Hungarian SINR& RSRP& $1.104564$  \\
\hline
Hungarian SINR + SLNR& Hungarian SINR& SLNR& $1.146535$  \\
\hline
Hungarian SINR + RSRQ& Hungarian SINR& Directed RSRQ& $0.949635$  \\
\hline
RZF Full CSI (RSRP assoc.)& Hungarian RSRP& Full CSI& $1.491904$  \\
\hline
RZF Full CSI (SINR assoc.)& Hungarian SINR& Full CSI& $1.491501$  \\
\hline
\end{tabular}
\label{tab2}
\end{center}
\end{table}

\section{Conclusions}\label{sec:conclusion}
This paper examines metric-driven beam management in 5G NR in three dimensions. First, RSRP, scalar RSRQ, and pre-equalization SINR are shown to be equivalent beam selection criteria when the interference estimate does not vary across beam indices, establishing a formal condition under which alternative metrics provide no gain over RSRP. Scalar RSRQ uses a beam-independent interference denominator (RSSI) and is therefore equivalent to RSRP as a metric for beam selection, whereas directional RSRQ aims to maximize RSRQ by minimizing RSSI (potentially reducing interference but not necessarily), breaking this equivalence. Second, two counterintuitive results emerge from a 3GPP-compliant 7-cell UMi simulation: directional RSRQ is the worst-performing strategy in symmetric interference as aggressive spatial avoidance selects geometrically poor beams, and sequential UE-gNB assignment outperforms power-optimal Hungarian strategies, indicating that preserving sector geometry benefits spatial precoding more than maximizing aggregate received power. Third, matched RZF full-CSI upper bounds confirm that the residual performance gap is attributable to codebook quantization rather than metric imperfection, pointing to adaptive or hierarchical codebook design as a mechanism to enhance performance. Non-uniform interference scenarios such as directional jamming or asymmetric cell loading, and a learned weighted multi-metric selection criterion, are identified as primary future directions.
\section*{Acknowledgment}
The system model and results were all generated on NVIDIA's Sionna software \cite{sionna}. The author used Claude (Anthropic) \cite{claude2024} and Grammarly \cite{grammarly} to assist with editing and condensing portions of the manuscript text.

\bibliographystyle{IEEEtran}

\bibliography{references} 
\nocite{*}

\end{document}